\documentclass[
aps,prl,twocolumn,groupedaddress
]{revtex4-1}

\usepackage{graphicx}
\usepackage{dcolumn}
\usepackage{bm}
\usepackage{natbib}
\usepackage{placeins}
\usepackage{braket,mleftright}
\usepackage{amsmath}
\usepackage{xcolor}

\begin{document}

\preprint{APS/123-QED}

\title{Temperature Dependence of Joule heating in Zigzag Graphene Nanoribbon}

\author{Yanbiao Chu, Cemal Basaran}
\email{cjb@buffalo.edu}
\affiliation{Electronic Packaging Laboratory, University at Buffalo, SUNY, Buffalo, NY, 14260, USA}

\date{\today}

\begin{abstract}
Using full-band electron and phonon dispersion relation, we investigate the temperature dependence of Joule heating in Zigzag Graphene Nanoribbons under high-field. At different temperatures of 300 K, 600 K, and 900 K, the Joule heating always increases linearly with time or the power is a constant. Although the scattering rates at 900 K and 600 K are 113 and 55 times higher than that of 300 K, the Joule heating power of 900 K and 600 K are just 12.6 and 6.7 timers higher than that of 300 K.

\begin{description}

\item[PACS numbers]
76.67.-n, 78.40.-q 

\end{description}

\end{abstract}

\pacs{Valid PACS appear here}
\keywords{}
\maketitle



The excellent electrical, thermal, and mechanical properties of graphene nanoribbons hold many promising application in electronics \cite{geim2007rise,han2007energy,li2008chemically,geim2009graphene,schwierz2010graphene}. To design energy-efficient circuits and energy-conversion systems, it is of great importance to understand energy dissipation and transport in nanoscale structures. The dissipated electric power can then raise the operating temperature to a point where thermal management becomes critical. The high current-carrying capacity is also critical for reliability, as many research works demonstrate Joule heating as the main failure mechanism \cite{PhysRevLett.100.206803,PhysRevLett.106.256801,murali2009breakdown}.

 The straightforward measurement method is to probe limiting current density by measuring I-V until devices break down. For solution-deposited GNRs of sub-10nm width \cite{PhysRevLett.100.206803}, the limiting current density is around $2mA/\mu m$. As bilayer graphene was used here, the current density is equivalent to $10^8A/cm^2$. Another experiment based on solution deposited GNRs was carried out by Liao et. al. \cite{PhysRevLett.106.256801}, they found that the maximum current density is limited by self-heating. For GNRs with ~15 nm wide, the current can reach $3mA/\mu m$. These values are obtained for GNRs without perfect edge due to possibly mixed edge shape and dangling bonds. Thus, edge scattering plays an important role in narrow GNRs. More promising results can be obtained with better fabrication methods available. For exfoliated GNRs up to five layers, Murali et. al. \cite{murali2009breakdown} demonstrated a limit current density of 1.2 - 2.8$\times 10^8A/cm^2$. And the breakdown current density is found to have a reciprocal relationship to GNR resistivity, which fit points to Joule heating as the likely mechanism of breakdown.
 
To assess the intrinsic current carrying ability of GNRs, however, theoretical analysis must be sought to study the carrier transport and scattering mechanism in GNRs. In this study, we investigate the temperature dependence of Joule heating in Zigzag Graphene Nanoribbons under high-field, using Ensemble Monte Carlo simulations with full-band electron and phonon dispersion relation. 

\begin{figure}[h]
\centering
\includegraphics[scale=0.45]{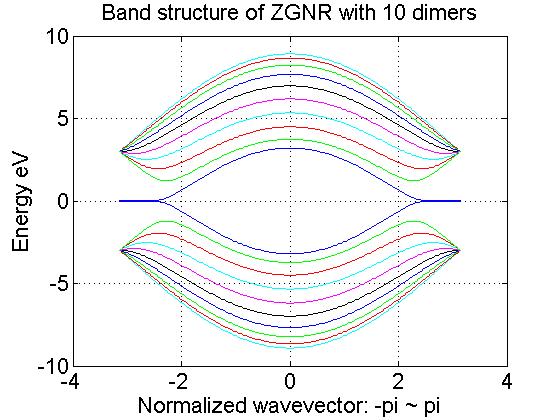}
\caption{\label{fig:ZGNR} Electron bands structure of ZGNR with 10 dimers. For the conduction bands with energy above 0, the sub band number $i$ is assigned as 1 to 10 from the highest band to the lowest band.}
\end{figure}

Within the formulation of tight-binding method, the energy bands of ZGNR with N dimers can be solved analytically as \cite{1468-6996-11-5-054504},
\begin{equation}
\begin{split}
E_i=&\left\{g_k\cos(\xi_i(N+1))  +\cos(\xi_i N)\right\}\\
&+i\left\{g_k\sin(\xi_i(N+1)) + \sin(\xi_i N)\right\},
\end{split}
\end{equation}
where $g_k=2\cos(k/2)$ and $k$ in the longitudinal wave number. The sub band $i$ is labelled by the quantization parameter $\xi_i$ in the transveral direction, which is determined by, 
\begin{equation}
F(\xi_i,N)\equiv g_k\sin(\xi_i(N+1)) + \sin(\xi_i N)=0.
\end{equation}
The reason is that the energy $E_i$ should always be a real number, thus the imaginary part should be zero. 
For ZGNR with even dimers, each sub band $i$ can be shown to have a parity of even or odd, which is identical to the even or odd of $i$ [ref: APL paper]. For ZGNR with 10 dimers, the band structure is shown in Fig. \ref{fig:ZGNR}.

The band gap in Fig.~\ref{fig:ZGNR} is 0, which is intrinsic to the tight-binding model and is always 0 for ZGNRs with any dimers. Based on these results, ZGNRs are said to be metallic and proposed as a promising replacement for conventional interconnects in electronics \cite{hod2007enhanced, xu2009modeling, naeemi2009compact}. But according to first principle calculations \cite{son2006half}, ZGNRs have a band gap depending on the width of nanoribbon and the magnitude is on the order of 0.2 eV. However, the Fermi level of ZGNR can be easily lifted to overcome this band gap. In this study, the Fermi level is set at the bottom of conduction band and all valence bands are fully occupied. Thus, the conduction bands obtained by tight-binding method works well as the first approximation.

\begin{figure}[h]
\hbox{\hspace{0cm}\includegraphics[scale=0.45]{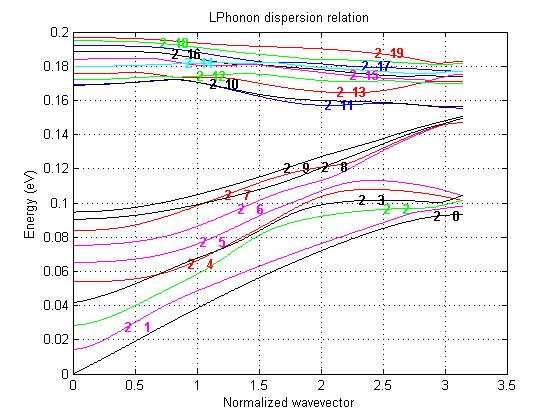}}
\caption{\label{fig:Phonon} Longitudinal phonon modes for ZGNRs with 10 dimers. For each band, the first number means the vibration direction, here 2 correspondes to the longitudinal direction. The second number means the number of nodes $n$ in each mode. Nodal number $n$ bigger than 9 correspones to optical modes, as the real value should be taken as $n-10$.}
\end{figure}

The phonon dispersion relation of ZGNR is obtained by Force Constant Methods \cite{saito1998physical}, as shown in Fig. \ref{fig:Phonon}. Similar to the vibration of membranes with two free edges, the modal shapes have the forms of sine functions (odd) or cosine functions (even). Approximately, the wave length $\lambda$ and wave number $\eta$ of mode $p$ can be expressed as
\begin{equation}
\lambda_p=\frac{2W}{n}, \qquad \eta_p=\frac{2\pi}{\lambda}=\frac{n}{W}\pi,
\end{equation}
where $W$ is the width of ZGNR and $n$ is the number of nodes in the width direction. The modal shape and parity were also verified by first principle calculations \cite{PhysRevB.77.054302,PhysRevB.80.155418}. Here, we only consider the longitudinal modes as their deformation potential are quite big than those of transverse mode and out-of-plane modes \cite{apl88/14/10.1063/1.2191420}.

Based on the preceding discussion, both electron and phonon have the parity. Due to geometricall mirror symmetry for lattice of ZGNR with even dimers, the parity should be conserved in the electron and phonon interaction process [cite APL]. Other than this, the electron can be scattered by a phonon to any sub bands without violating parity conservation, this the so called transverse momentum conservation uncertainty \cite{Betti2007APL}. As usual, the selection rule also includes the conservation of energy and the conservation of longitudinal momentum. By Fermi's Golden Rule, the scattering rates can be calculated as
\begin{equation}
\Gamma(k,k')=\frac{2\pi}{\hbar}|\mathcal{M}(k,k')|^2 ,
\end{equation}
where $\mathcal{M}(k,k')$ is the matrix element for an electron gets scattered by a specific phonon mode from intial state $k$ to final state $k'$. 

For the three-particle process here, the parity conservation is equivalent to: 
\begin{equation}
Parity(p_1)\times Parity(p_2)\times Parity(p_3)=even,
\end{equation}
where $p_i$ represents either a phonon or electron. In other words, for the scattering process with electron jumping from even state to even state or odd state to odd state, the parity of the involved phonon can only be even. Similarly, only odd phonons can scatter electrons from even state to odd state or odd state to even state. 

Therefore, the matrix element $\mathcal{M}({k,k'})$ is always even about the dimer index $d$ and can be written as
\begin{equation}
\begin{split}
 &\mathcal{M}({k,k'})= \delta_{g,k+q-k'}2\sum_{d=1}^{N/2}  i \sqrt{\frac{\hbar n_{q,p}}{2N_c m \nu_q}}D(1+e^{i\theta_{kk'}})    \\
&\times \frac{\psi_{d}}{N_e} \frac{\psi'_d}{N'_e}
\begin{cases}
    \cos(\frac{\eta(N+1-2d)}{2})/N_{ph}^e, & \text{even phonon}.\\
    \sin(\frac{\eta(N+1-2d)}{2})/N_{ph}^o, & \text{odd phonon}.
  \end{cases}
\end{split}
\end{equation}
Here, $N_c$, $m$, and $n_{q,p}$ are the number of unit cell  in the system, mass of carbon atom, and number of phonons respectively; the Kronecker delta $\delta_{g,(k-k'-q)}$ describes the conservation of momentum in the longitudinal direction; $\psi_{d}=\sin{\xi(N+1-2d)/2}$ is the magnitude of electron wave and is normalized by $N_e=\sum_d \sin^2{\xi(N+1-2d)/2}$; similarly, $\cos{\eta(N+1-2d)/2}$ and $N_{ph}^e=\sum_d\cos^2{\eta(N+1-2d)/2}$ are magnitude and normalization constant for even phonons, while $\sin{\eta(N+1-2d)/2}$ and $N_{ph}^o\sum_d\sin^2{\eta(N+1-2d)/2}$ are magnitude and normalization constant for odd phonons; the deformation potential $D$ for optical phonon is $1.4 \times 10^9$eV/cm \cite{PhysRevB.78.205403}, and the deformation potential for acoustic phonon is 16 eV \cite{apl/90/6/10.1063/1.2437127}.

\begin{figure}[h]
\centering
\includegraphics[scale=0.2]{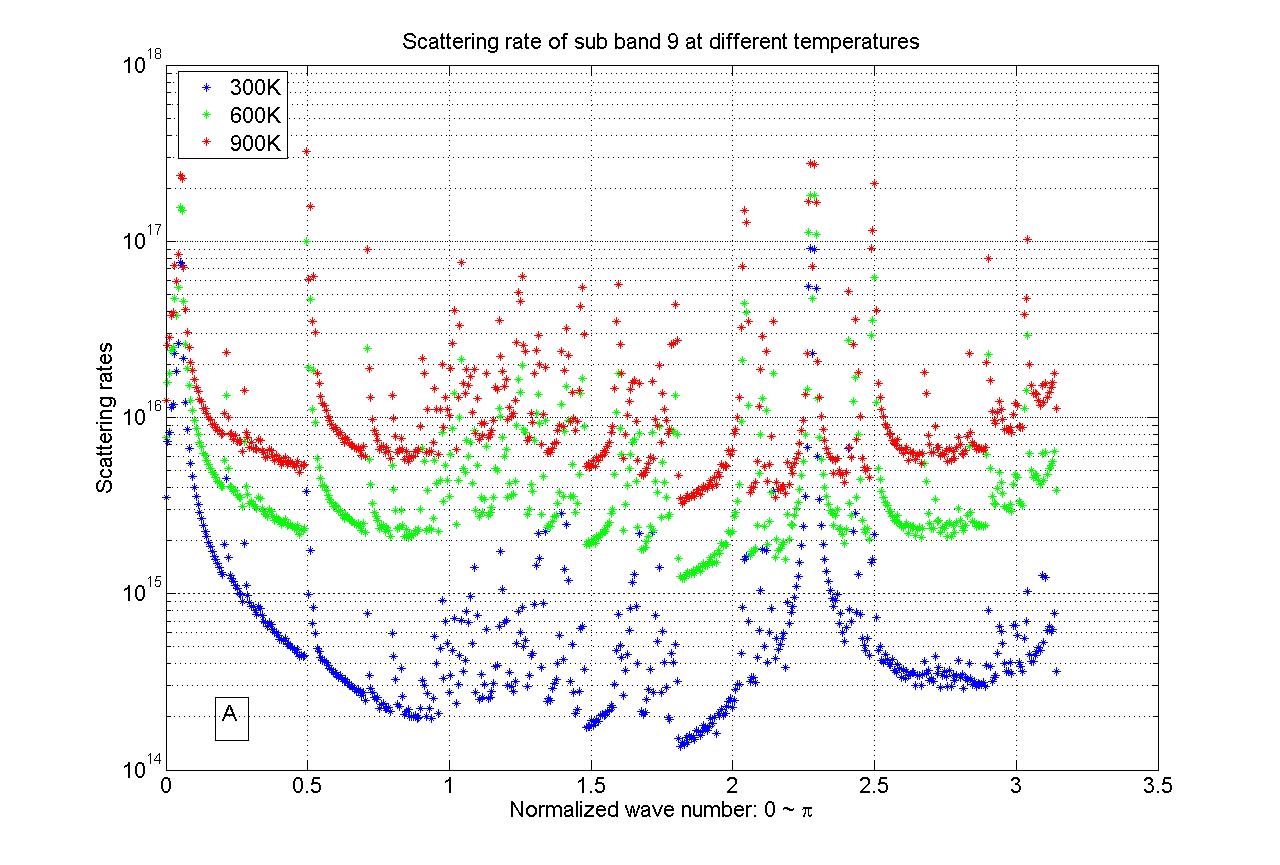}
\includegraphics[scale=0.2]{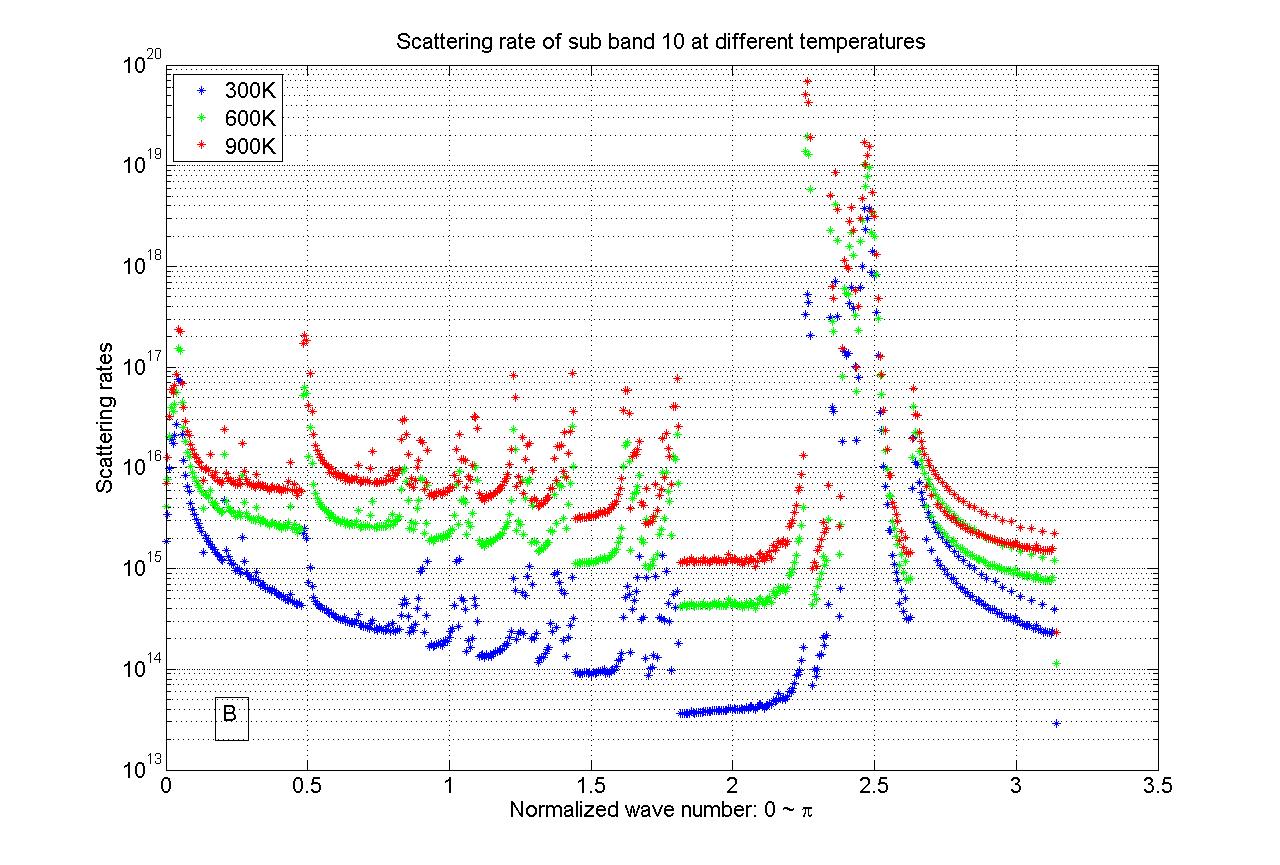}
\caption{\label{fig:Scattering} Scattering rates for ZGNR at temperature 300K, 600K, and 900K. Only results for electron sub band 9 (A) and electron sub band 10 (B) are shown. The rate for each electron state is a sum of all available scattering mechanisms.}
\end{figure}

The scattering rates for ZGNR with 10 dimers are demonstated in Fig. \ref{fig:Scattering}. Since the Fermi level is set at the bottom of conduction band, most electron resides at the lowest conduction band (electron sub band 10), only the scattering rates of sub band 9 and sub band 10 are presented. Generally, the scattering rates at room temperature (300 K) is orders higher than the values for Carbon Nanotube. One reason is that the absence of the periodical boundary condition in the restricted direction of ZGNR, which removes the selection rule of subband number conservation. Therefore, each electron can interact with all 10 branches of phonons and justifies the difference between ZGNR and CNT.

Another reason for the much higher scattering rates is due to the formulation. According to Betti et. al. \cite{Betti2007APL}, the term related to mass in scattering rates is $1/\rho W$, where $W$ is the width of nanoribbon and $\rho$ is the 2D graphene mass density. However, the expression for this term is $a/m$ with $a$ as the unit length of ZGNR. Obviously both of them has a dimension of length density. But in our formulation the term is a constant while in the formualtion of Betti et. al. it denpends on the width $W$. For wider nanoribbons $W$ could be tens of $a$ and result in a scattering rates tenth of our results.

As shown in Fig. \ref{fig:Scattering}, scattering rates at the same temperature are quite different even for different electron states (with different wave number) of the same sub band. To describe the free flight time between scattering, however, we need to assume a nominal scattering rate $\Gamma_0$ for all different electron states \cite{ lundstrom2009fundamentals}. The parameter $\Gamma_0$ is very important as it determines the size of reasonable time step in the Ensemble Monte Carlo (EMC), which in turns determines the efficiency of EMC simulation. According to Fig. \ref{fig:Scattering}, the highest scattering rates for electrons in sub band 10 are on the order of $10^{19}$. Therefore, the time step should be set on the order of $10^{-20}$ s, which is unreasonable small. Another problem is that the highest and lowest scattering rates for sub band 10 are of 5 oders different even at the same temperatures. In the scattering mechanisms scattering process, this implies that the self scattering probability is $99.999\%$ after the free flight. In other words, the physical probability for electron get scattered is only 0.001\%, which makes the simulation extremely inefficient.

We should make two observations here. First, the electron states with highest scattering rates are those with wave numbers around 2.5 (normalized by 1/a). In tight-binding model for ZGNR with 10 dimers, all electron states in sub band 10 with wave numbers larger than 2.13 are the almost flat edge states with extremely density of state. According to first principle calculations, the segment of band 10 corresponding to edge states is not flat. Second, the detailed inspection of the final states for all electrons in this range shown that, the involved phonons have almost zero energy. As we are only interested in the energy transfer between phonons and electrons, the contribution of these scattering events is very week. Therefore, we can normalize these scattering rates to $10^{15}$ to expedite our simulation.

\begin{figure}[h]
\centering
\includegraphics[scale=0.25]{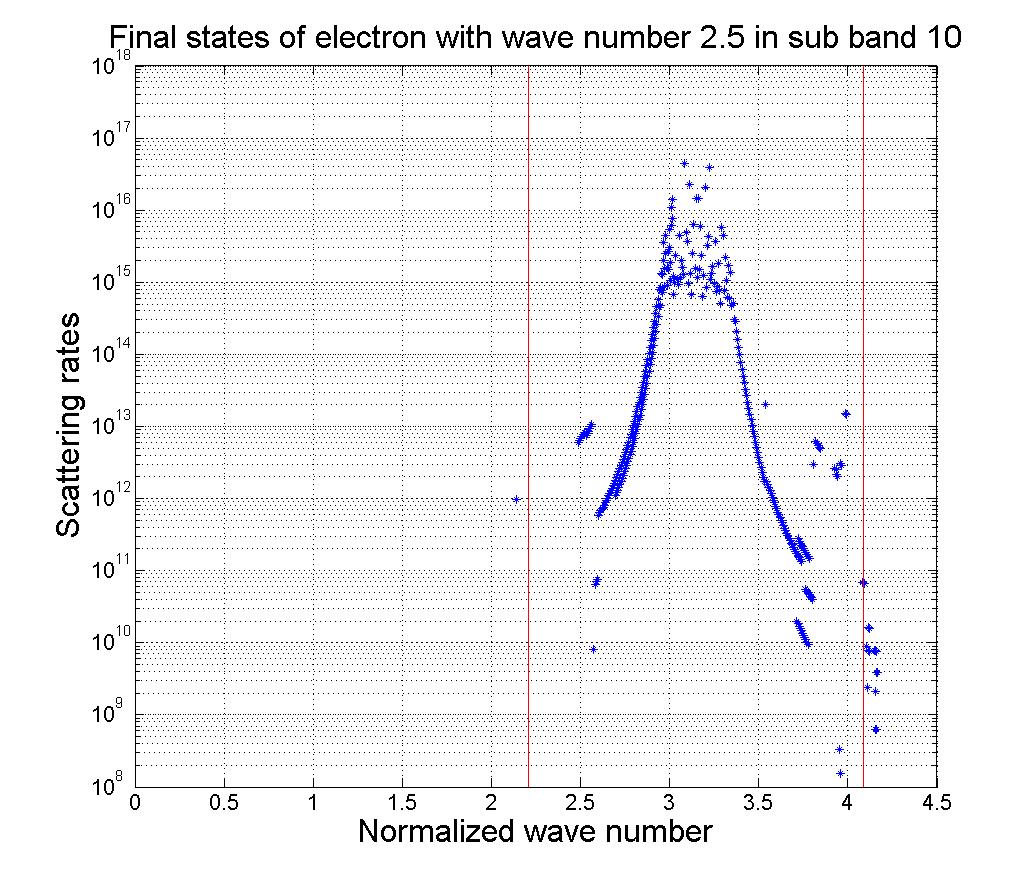}
\caption{\label{fig:Final} Distribution of final states after scattering for electron with wave number 2.5 in sub band 10 at 300 K. As showin in Fig. \ref{fig:ZGNR}, the energy of states between two red lines are almost zero, and the energy difference between edge states in band 10 with all other nine sub bands is bigger than 1 eV. Given that the highest energy of all phonon modes is about 0.2 eV, those edge states can only get scattered within sub band 10.  }
\end{figure}

\begin{figure}[h]
\centering
\includegraphics[scale=0.2]{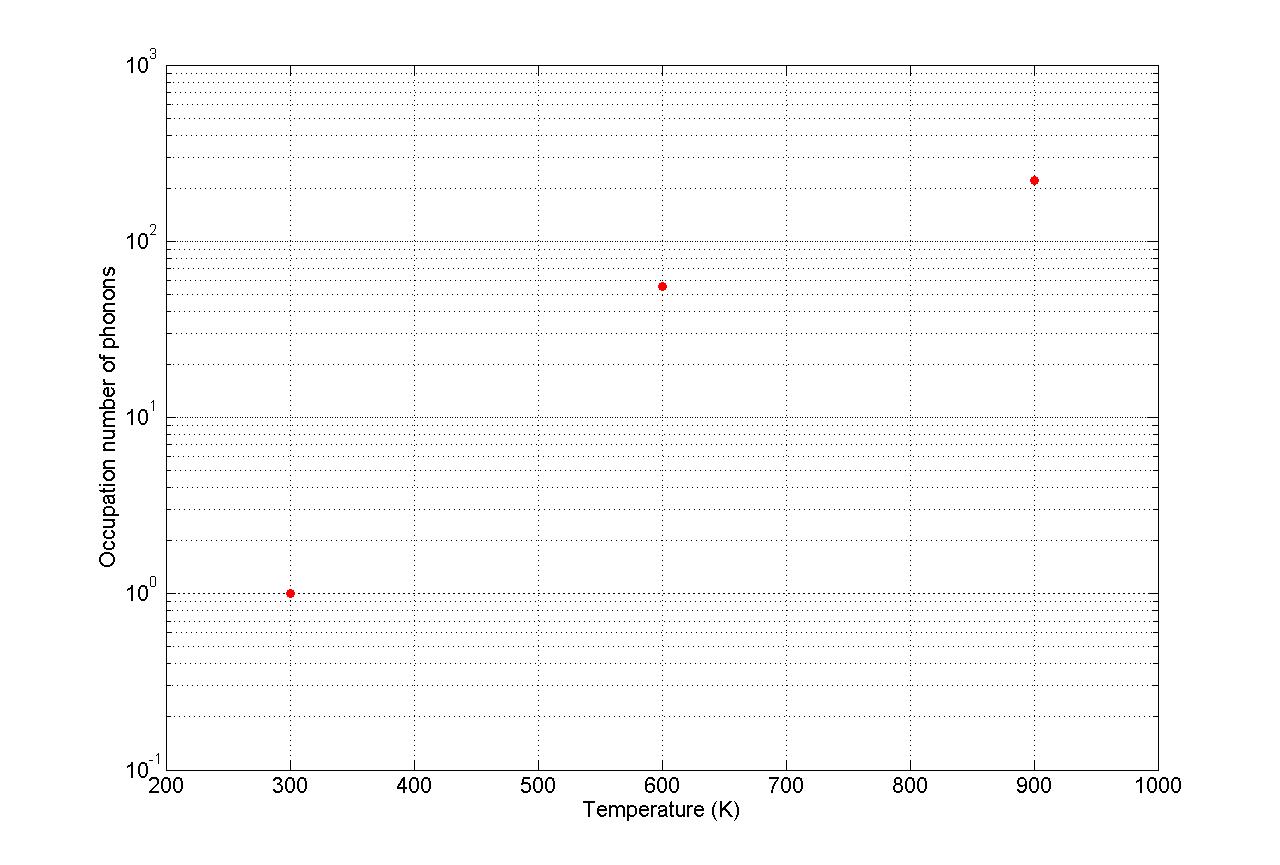}
\caption{\label{fig:Occupation} Demonstration of occupation numbers for an phonon of 0.2 eV, normalized by the occupation number at 300 K.}
\end{figure}

In the formulation of scattering rates, the temperature denpendence only comes from the phonon occupation number $n_{q,p}$. As one kind of Bosons, $n_{q,p}$ of phonons follow the Bose-Einstein distribution,
\begin{equation}
n_{q,p} = \frac{1}{exp(n_{q,p}\hbar \nu_{q,p}/k_B T)-1},
\end{equation}
wherer $T$ is the temperature. Take optical phonons as an example, whose energy can be assumed as constant with the value of 0.2 eV. The occupation numbers at different temperature (300 K, 600 K, and 900 K) are shown in Fig. \ref{fig:Occupation}. The occupation number at 600 K is about 55 times higher than the occupation number at 300 k, while that at 900 K is 113 times higher. This explains the magnitudes of scattering rates at different temperatures.

In the EMC simulations, the first Brillouin zone for both electron and phonon is discretized by 1000 points. According the quantization scheme of Bloch's Theorem, this implies that we are studying a system of the 1000 unit cells long in the real space.  For a specific temperature and Fermi level, the calculated electron occupation number $N_e$ in reciprocal space is equivalent to $N_e/1000$ electrons per unit cell in real space. Correspondingly, the linear electron density in real space is $N_e/1000/a$. With the Fermi level set as 0, the occupation number $N_e$ of 300 K, 600 K, and 900 K in reciprocal space are 54, 59, and 63 respectively. 
According to the preceding discussion of renormalized scattering rates, the time step for temperature at 300 K, 600 K, and 900K are set as $8\times 10^{-3}$ fs, $6\times 10^{-3}$ fs, and $5\times 10^{-3}$ fs respectively. Each simulation runs for $10^4$ fs. And the high electrical field in this study is set as 20 kV/cm. 

Quantum mechanically, Joule heating power is the energy transfered from electrons to phonons and can be calculated as \cite{jap/112/10/10.1063/1.4766901},
\begin{equation}
w=\frac{\sum_{m,i}\int{(E_{k',i'}^m - E_{k,i})S_{k',i';k,i}^m f_{k,i}(1-f_{k',i'})dk}}{\pi}
\end{equation}
where $E_{k',i'}^m -E_{k,i}$ is the energy transferred during the scattering event from an initial state $(k,i)$ to a final state $(k',i')$ by scattering mechanism $m$ and the corresponding scattering rate is $S_{k',i';k,i}^m$; $f_{k,i}$ is the occupation probability of the state $(k,i)$, while $(1-f_{k',i'})$ is the probability that state $(k',i')$ is unoccupied.

\begin{figure}[h]
\centering
\includegraphics[scale=0.25]{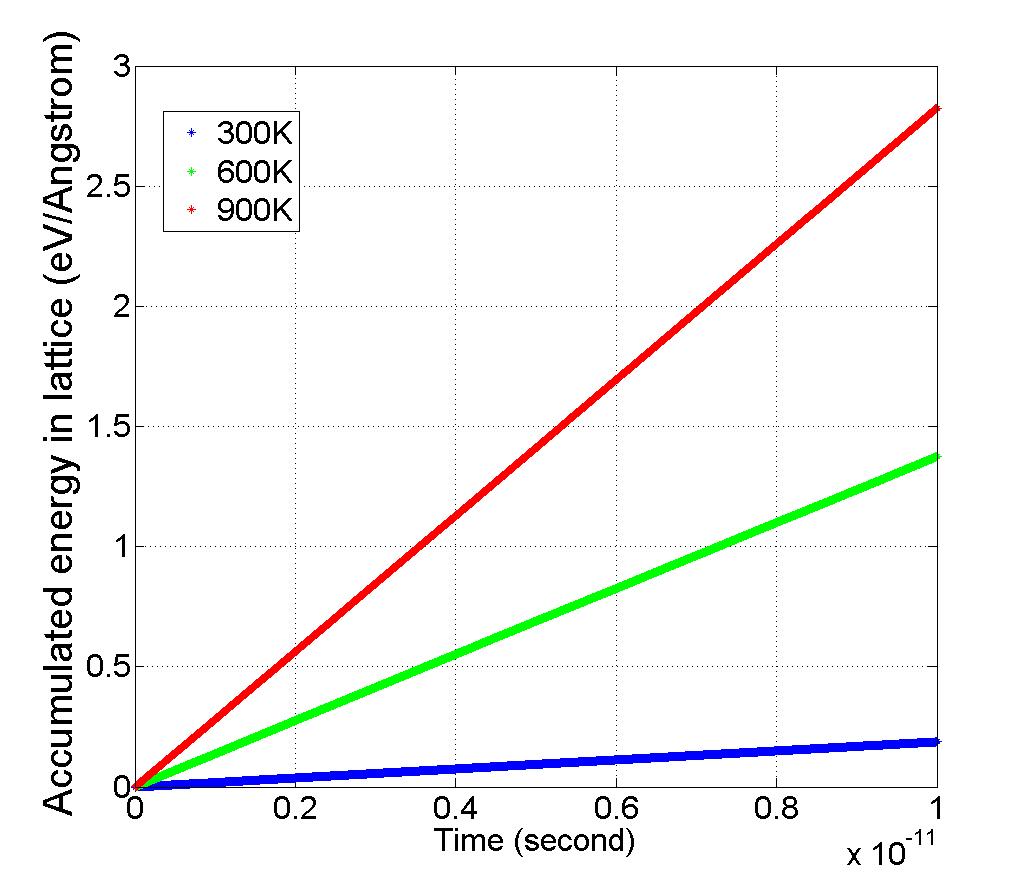}
\caption{\label{fig:Joule} Joule heating at different temperatures.}
\end{figure}


The results of EMC simulations are demonstrated in Fig. \ref{fig:Joule}.  It shows that the accumulated energy transfered from electrons to lattice at different temperatures always increases linearly with time. Or the Joule heating power is a constant. Although the scattering rates at 900 K and 600 K are 113 and 55 times higher than that of 300 K, the Joule heating power of 900 K and 600 K are just 12.6 and 6.7 timers higher than that of 300 K.

We would like to recognize the contribution of Dr. Xuedong Hu of the Physics Department at the University at Buffalo for our insightful discussions. We also gratefully acknowledge the financial support received from the US Navy Office of Naval Research Advanced Electrical Power Systems program, under the direction of Dr. Peter Chu. 

\bibliographystyle{apsrev4-1}
\bibliography{Joule_heating_in_ZGNR}

\end{document}